\documentclass[review]{elsarticle}

\usepackage{lineno,hyperref}
\usepackage{amssymb}
\usepackage{mathrsfs}
\usepackage{amsmath}

\newtheorem{lem}{Lema}
\newtheorem{corol}{Corolario}
\usepackage{epsfig}
\usepackage{graphicx}
\graphicspath{{graficos/}}
\usepackage{float}
\usepackage{}
\modulolinenumbers[5]

\begin{document}

\begin{frontmatter}

\title{New Lagrange multipliers for the time fractional Burgers' equation}

\author{A. R. G\'omez Plata\fnref{myfootnotee}}
\address{Imecc, Unicamp, 13083-859,  Campinas, SP\\Umng, Bogot\'a, COL.}
\fntext[myfootnote]{email: adrian.gomez@unimilitar.edu.co, capelas@ime.unicamp.br}

\author{E. Capelas de Oliveira\fnref{myfootnote }}
\address{ Imecc, Unicamp, 13083-859,  Campinas, SP}



\begin{abstract}
Using the fractional derivative, considered in the Caputo sense, we study  an analytical technique associated with the variational iteration method for the  fractional  generalized $\alpha$-time  Burgers' equation with $\alpha>0$ and obtain approximate   solutions in particular cases  $0<\alpha\leq 1$ and  $1<\alpha\leq 2$. 

\end{abstract}

\begin{keyword}
\texttt  Caputo derivative\sep  variational iteration method\sep Lagrange multipliers\sep Burgers' equation\sep Laplace transform.
\end{keyword}

\end{frontmatter}

\section{Introduction}

The fractional calculus (FC) is a very important tool associated with several problems which appear in physics, engineering an other sciences \cite{podlubny,gorenflo,Debnath,Yusufoglu}. As an important example we mention  diffusion processes, particularly  to the fractional partial Burgers' equation (FPBE) appearing in the traffic flow and gas  dynamics  \cite{Wang}. On the order hand, the variational iteration method (VIM) is a relatively  new approaches to provide an analytical approximation to linear and nonlinear problems \cite{Odibat,Bal,Momami}. Those authors  consider the VIM  applied to the time-fractional partial Burgers' equation.\\ Here we are interested in  the VIM associated with the so-called generalized FPBE 

\begin{equation}
{}_{c}{\sf D}_t^{\alpha} [u]= u_{xx} +Au^p u_{x},\quad u=u(x,t),
\end{equation}
\\
with ${\sf D}=\frac{\partial}{\partial t}, 0<\alpha\leq 1$ and  $1<\alpha\leq 2, A\in\mathbb{R}$ and $p>0$.\\

\noindent
Some particular results appearing in the literature, are recovered.

\noindent
The paper is organized as follows: in section 2, preliminaries  will be presented, particularly, a short review involving the FC, specifically the Caputo derivative  and its respective  Laplace transform, after of VIM it is analysed, in particular for nonlinear fractional partial differential equation. In section 3 we will present the VIM and the Burgers' equation and provide a lemma, with the proof involving the general case and showing some particular cases of the parameters  presented  as examples. In the section 4 we will discuss the  approximate  solutions for the generalized FPBE, recovering the results involving the classical Burgers' equation, presenting several examples with the respective graphics. Concluding remarks close the paper.

\section{Preliminaries}

In this section we will present  definitions and results that we use in the paper, a short review of FC and the VIM for the linear and nonlinear equations.

\subsection{Fractional Calculus}
First of all, we introduce the Riemann-Liouville (RL) fractional integral, considered 
in the left, only \cite{Diethel}. Let $n$ be a positive integer and $\alpha \in \mathbb{C}$ such that
Re$(\alpha) > 0$, we define the RL fractional integral by means of

\begin{equation}
{\sf I}_t^{\alpha} [f(t)] := \frac{1}{\Gamma(n - \alpha)} \int_0^t 
f(\tau){(t-\tau)^{\alpha + 1 - n}}{\mbox{d}}\tau\ ,\quad n-1 < \alpha < n.
\end{equation}

\noindent
The Caputo derivatives has been used by many authors in several physical applications \cite{Mainar,Silva,Oliveira,Figueiredo,Diethelm,Liu,Wazwaz}. 
One reason for this choice is the fact that the initial conditions associated with the fractional differential equation
are usually expressed in terms of integer order derivatives. Let $n$ be a positive 
integer and $\alpha \in \mathbb{C}$ such that Re$(\alpha) > 0$. We introduce the fractional 
derivative of order $\alpha$ in the Caputo sense, denoted by ${}_{c}{\sf D}_t^{\alpha}[f(t)]$, 
by means of the integral
\begin{equation}
{}_{c}{\sf D}_t^{\alpha} [f(t)] = \left\{
\begin{array}{l}
\displaystyle \frac{1}{\Gamma(n - \alpha)} \int_0^t \frac{f^{(n)}(\tau)}{(t-\tau)^{\alpha+ 1 - n}}{\mbox{d}}\tau\,,
\quad n-1 < \alpha < n,\\
\\
\displaystyle {\sf D}^n f(t) \equiv \frac{{\mbox{d}}^n}{{\mbox{d}}t^n} f(t)\,, \quad \alpha = n.
\end{array} \right.
\end{equation}
The relation involving the Caputo derivatives and the RL fractional integral is given by
\begin{equation}
{}_{c}{\sf D}_t^{\alpha}[f(t)] = {\sf I}_t^{n-\alpha} {\cdot} _{c}{\sf D}_t^n f(t)
\end{equation}
with the $n - 1 < \alpha < n$.

\noindent
As we have already said, the Laplace transform methodology is an efficient tool to discuss a fractional 
differential equation. Then, we introduce the Laplace transform of the derivative in the Caputo sense. Denoting 
by $\mathscr{L}$ the Laplace integral operator, we can write the Laplace transform of the 
Caputo derivatives as follows
\begin{equation}
(\mathscr{L}[{}_{c}{\sf D}_t^{\alpha} f(t)])(s) = s^{\alpha}(\mathscr{L}[f(t)])(s) - \sum_{k=0}^{n-1} 
s^{\alpha - k - 1} ({\sf D}^k f)(0^{+})\,,
\end{equation}
with $s$ is the parameter of the Laplace transform.

\noindent
As we have mentioned above, this expression shows that the Laplace transform of the fractional 
derivative in the Caputo sense involves only the derivative of integer order evaluated in $t=0^{+}$, 
conversely the corresponding RL derivative. In our particular problems, as we will be 
seen in the sequence, we take the parameter $\alpha$ as a real number such that $0 < \alpha \leq 1$ in 
problems involving (anomalous) diffusion and $1 < \alpha \leq 2$ in problems associated with 
wave propagation.

\noindent
To close this subsection, as an example, we consider the fractional integral and the fractional 
derivative of a power function $t^{\lambda}$, with $\lambda$  a real parameter. For the fractional 
integral we have
\begin{equation}
{\sf I}_t^{\alpha} [t^{\lambda}] = \frac{\Gamma(\lambda + 1)}{\Gamma(\lambda + 1 + \alpha)} t^{\lambda + \alpha}
\end{equation}
with $\alpha \geq 0$, $\lambda > -1$ and $t > 0$. On the other hand, for the fractional derivative we have
\begin{equation}
{}_{c}{\sf D}_t^{\alpha}  [t^{\lambda}] = \frac{\Gamma(\lambda + 1)}{\Gamma(\lambda + 1 - \alpha)} t^{\lambda - \alpha}
\end{equation}
with $\alpha > 0$, $\lambda > -1$ and $t > 0$.

\subsection{Variational iteration method }

The VIM \cite{He1,He2,He3} was extended to fractional differential equations and has been one of  the methods frequently used. Classical and fractional differential equations are studied using  VIM \cite{He2,He3}. On the order hand classical and  fractional partial differential equations are studied in \cite{Mainardi,Waz1,Waz2,Ozer,Sheng,Ruiz,Hayat,Shah,Chen} in particular nonlinear dynamics for local fractional Burgers' equation arising in fractal flow is discussed in \cite{Tenreiro}.

\noindent
Here we consider a more general fractional differential equation 
\begin{center}
\quad ${}_{c}{\sf D}_t^{\alpha} [u]+R[u]+N[u]=f(t)$,
\end{center}
\noindent
where ${}_{c}{\sf D}_t^{\alpha} [u]$ is the Caputo derivative, $R[u]$ is a linear term, $N[u]$ is a nonlinear one and $f(t)$ is a function associated with the non homogeneous term. Odibat and Momami in \cite{odibat} applied the VIM to the above equation and suggested a variational iteration formula 

\begin{equation*}
\left\{
\begin{array}{l}
\displaystyle u_{n+1}=u_n+\int_0^t\lambda(t,\tau)(_{c}{\sf D}_t^{\alpha}u_n+R[u]+N[u]-f(\tau)){\mbox{d}}\tau,\qquad 0<t\\
       \\
\displaystyle \lambda(t,\tau)=-1,\qquad 0< \alpha\leq 1\\
\\
\displaystyle \lambda(t,\tau)=\tau-t,\qquad 1< \alpha\leq 2.
\end{array} \right.
\end{equation*}

\noindent
$\lambda(t,\tau)$ are known as the Lagrange multipliers associated with the variational iteration formula, this multipliers are evaluated with the general theory of Lagrange multipliers \cite{Inokuti}.

\section{VIM and fractional Burgers' equation}
In this section, we will present the VIM applied to the Burgers' equation:
\begin{equation}
{}_{c}{\sf D}_t^{\alpha} [u]= u_{xx} +Au^p u_{x},\quad u=u(x,t)
\end{equation}
with ${\sf D}=\dfrac{\partial}{\partial t}, 0<\alpha\leq 1$ and  $1<\alpha\leq 2, \quad A\in\mathbb{R}$ and $p>0$.\\\\

\noindent 
The so-called correction functional  for Eq.(8) is 

\begin{equation}
u_{k+1}(x,t)= u_k(x,t)+\frac{1}{\Gamma (\beta)}\int_{0}^{t}(t-\tau)^{\beta-1}\quad\lambda(t)\left(\dfrac{\partial^{\alpha}}{\partial t^{\alpha}}u_k-\dfrac{\partial^{2}}{\partial x^{2}}\bar{u}_{k}-A\bar{u}_{k}^{p}\dfrac{\partial}{\partial x}\bar{u}_k \right) {\mbox{d}}\tau,
\end{equation}

\noindent
where $\beta=\alpha+1-m$ and $m-1<\alpha\leq{m}$.\\
\noindent
The Eq.(9)  can be approximately expressed by means of
\begin{equation}
u_{k+1}(x,t)=u_k(x,t)+\int_{0}^{t}\lambda(t) \left(\dfrac{\partial^{m}}{\partial \tau^{m}}u_k-\dfrac{\partial^{2}}{\partial x^{2}}\bar{u}_{k}-A\bar{u}_{k}^{p}\dfrac{\partial}{\partial x}\bar{u}_{k}  \right){\mbox{d}}\tau.
\end{equation}
If  we use integration by parts and remembering that the stationary term in the functional  $\delta\bar{u}_{k}=0$, we get three cases as follow:\\

\noindent
a) For $m=1$ we have $0 < \alpha \leq 1$ and the correction functional  can be approximately by means of\\ 
\begin{equation*}
\delta u_{k+1}(x,t)=\delta u_{k}(x,t)+\delta \int_{0}^{t}\lambda(t)\left(\dfrac{\partial}{\partial t}u_k-\dfrac{\partial^{2}}{\partial x^{2}}\bar{u}_{k}-A\bar{u}_{k}^{p}\dfrac{\partial}{\partial x}\bar{u}_k\right){\mbox{d}}\tau.
\end{equation*}

\noindent
Thus,  we have the system

\begin{equation}
\begin{split}
1+\lambda(\tau)&=0\\
\lambda^{\prime}(\tau)&=0,\\
\end{split}
\end{equation}

\noindent
whose solution is $\lambda(\tau)=-1$. Then,  we obtain the following formula (note that, here we have $\beta=\alpha$) 

\begin{equation}
u_{k+1}(x,t)=u_{k}(x,t)-\frac{1}{\Gamma(\alpha)}\int_{0}^{t}(t-\tau)^{\alpha-1}\left(\dfrac{\partial^{\alpha}}{\partial t^{\alpha}}u_k-\dfrac{\partial^2}{\partial x^2} u_{k}-A u_{k}^{p}\dfrac{\partial}{\partial x}{u}_{k} \right){\mbox{d}}\tau.
\end{equation}

\noindent
b) For $ m =2$ we have $1 <\alpha\leq 2$ and the correction functional  can be approximately expressed by means of\\ 
\begin{equation*}
\delta u_{k+1}(x,t)=\delta u_{k}(x,t)+\delta \int_{0}^{t}\lambda(t)\left(\dfrac{\partial^2}{\partial t^2}u_k-\dfrac{\partial^2}{\partial x^2}\bar{u}_{k}-A\bar{u}_{k}^{p}\dfrac{\partial}{\partial x}\bar{u}_{k}\right) {\mbox{d}}\tau.
\end{equation*}\\

\noindent
Thus, we get the system:
\begin{equation}
\begin{split}
\lambda^{\prime\prime}(\tau)=0,\\
\lambda(\tau)_{\tau=t}=0,\\
1-\lambda^{\prime}(\tau)_{\tau=t}=0,
\end{split}
\end{equation}

\noindent
whose solution can be written as $\lambda(\tau)=\tau-t$. Then,  for $ m = 2 $ and $1 <\alpha\leq 2$, we obtain the following interaction formula with $\beta=\alpha-1$

\begin{equation}
u_{k+1}(x,t)=u_k(x,t)-\frac{\alpha-1}{\Gamma(\alpha)}\int_{0}^{t}(t-\tau)^{\alpha-1}\left(\dfrac{\partial^{\alpha}}{\partial t^{\alpha}}u_k-\dfrac{\partial^2}{\partial x^2} u_{k}-A u_{k}^{p}\dfrac{\partial}{\partial x}{u}_{k} \right){\mbox{d}}\tau.
\end{equation}

\noindent
c) News Lagrange multipliers with $0<t$ and $0<\alpha$. In this case, our main result,  we propose a lemma with its proof, recover the two precedent results and present an example. \\

\noindent
\begin{lem} If the correction functional of the{\rm{ Eq.(9)}} is given by the Riemann integration

\begin{equation}
u_{n+1}=u_{n}+\int_{0}^{t}\lambda(t,\tau)\left[ _c{\sf D}_t^{\alpha}(u_{n})-(\bar{u}_{n})_{xx}-A(\bar{u}_{n})^{p}(\bar{u}_{n})_{x}\right]{\mbox{d}}\tau,
\end{equation}

\noindent
with $0 <t$,  $0<\alpha$ and $(\bar{u}_{n})_{xx}$, $A(\bar{u}_{n})^p(\bar{u}_{n})_x$  are restrictions of the variations of the functional associated with {\rm{ Eq.(15)}}, then  the Lagrange multiplier is

\begin{equation*}
\lambda(t,\tau)=\frac{(-1)^{\alpha}(\tau-t)^{\alpha-1}}{\Gamma(\alpha)}.
\end{equation*}\\\\
\end{lem}

\textbf{Proof.} First of all, we transform Eq.(15) in its integral form and taking the Laplace  transform on both sides of  the new equation  
\begin{equation}
\mathscr {L}[u_{n+1}]=\mathscr {L}[u_n+{\sf I}_t^{\alpha}\lambda_{RL}(t,\tau)(_{c}{\sf D}_{\tau}^{\alpha}u_n-(\bar{u}_n)_{x x}-A(\bar{u}_n)^{p}(\bar{u}_n)_x)]
\end{equation}
where $\lambda_{RL}(t,\tau)$ is the Lagrange multiplier of the integral form for  Eq.(15).  We consider
\begin{equation}
{\sf I}_t^{\alpha}\lambda_{RL}(t,\tau)_{c}{\sf D}_t^{\alpha}u_n=\frac{1}{\Gamma(\alpha)}\int_{0}^t(t-\tau)^{\alpha-1}\lambda(t,\tau)_c {\sf D}_t^{\alpha}u_n(\tau){\mbox{d}}\tau.
\end{equation}
Thus, if  $\lambda_{RL}(t,\tau)=\lambda(t-\tau)$, Eq.(17) is the convolution of the function
\begin{equation}
a(t)=\frac{\lambda(t)t^{\alpha-1}}{\Gamma(\alpha)}
\end{equation}
\noindent
and  $_c {\sf D}_t^{\alpha}u_n(t)$.
The terms $(\bar{u}_n)_{x x}$ and $A(\bar{u}_n)^p(\bar{u}_n)x$ which are considered restrictions on variations, implying $\delta(\bar{u}_n)_{x x}=0$ and $\delta A(\bar{u}_n)^p(\bar{u}_n)x=0$.

\noindent
Using the variational functional associated with Eq.(16) and the Laplace transform of the Caputo derivative, we obtain

\begin{equation*}
\delta\mathscr {L}[u_{n+1}]=\delta\mathscr {L}[u_{n}]+\delta\mathscr {L}[{\sf I}_t^{\alpha}\lambda_{RL}(_c {\sf D}_t^{\alpha}u_n-(\bar{u}_n)_{x x}-A(\bar{u}_n)^p(\bar{u}_n)x)],\\
\end{equation*}

\begin{equation*}
=\delta\mathscr {L}[u_{n}]+\delta[\mathscr {L}[a(s)]s^{\alpha}\mathscr {L}[u_n(s)]]-\delta\sum_{k=0}^{m-1}u^k(0^+)s^{\alpha-1-k},\\
\end{equation*}

\begin{equation}
= (1+ \mathscr {L}[a(s)] s^{\alpha})\delta\mathscr {L}[u_n],
\end{equation}\\

\noindent
where, using the Euler-Lagrange equation associated with Eq.(16), we have
\begin{equation*}
1+\mathscr{L}[a(s)]s^{\alpha}=0,
\end{equation*}

\noindent
resulting

\begin{equation}
\mathscr{L}[a(s)]=-\frac{1}{s^\alpha}.
\end{equation}
 Performing the inverse Laplace transform of Eq.(20) we get
\begin{equation}
a(t)=-\frac{t^{\alpha-1}}{\Gamma(\alpha)}.
\end{equation}

\noindent
Comparing Eq.(18) and Eq.(21) we have $\lambda_{RL}(t,\tau)=-1$ and using Eq.(16) we can write

\begin{equation*}
\displaystyle
u_{n+1}=u_n-{\sf I}_t^{\alpha}[_c {\sf D}_t^{\alpha}u_n-(\bar{u}_n)_{xx}-A(\bar{u}_n)^p(\bar{u}_n)_{x}],\\
\end{equation*}
\begin{equation*}
\displaystyle
{u_{n+1}=u_n}-{\displaystyle{\int_{0}^{t}}}\displaystyle{\frac{(t-\tau)^{\alpha-1}}{\Gamma(\alpha)}} {{\left[_{c}{\sf D}_t^{\alpha}u_n-(\bar{u}_n)_{xx}-A(\bar{u}_n)^p(\bar{u}_n)_{x} \right]}} {\mbox{d}}\tau,\\\\
\end{equation*}
\begin{equation*}
\displaystyle
u_{n+1}=u_{n}+\int_{0}^{t}\frac{(-1)^{\alpha}(\tau-t)^{\alpha-1}}{\Gamma(\alpha)} \left[ _{c}{\sf D}_{t}^{\alpha}(u_{n})-(\bar{u}_{n})_{xx}-A(\bar{u}_{n})^{p}(\bar{u}_{n})_{x} \right]\mbox {d}\tau.
\end{equation*}

\noindent
Finally, using Eq.(17) we obtain the following general expression 
\begin{equation}
\lambda(t,\tau)=\frac{(-1)^{\alpha}(\tau-t)^{\alpha-1}}{\Gamma(\alpha)}.
\end{equation}

\noindent
In what follows, we recover the known results and discuss an example.

\begin{corol}
In {\rm{Eq.(22)}} if $\alpha=1$ then $\lambda(t,\tau)=-1$ and if $\alpha=2$ then $\lambda(t,\tau)=\tau-t$.\\
\end{corol}

\textbf{Proof.}
Direct substitution $\alpha=1$ ($\alpha=2$) in Eq.(22).\\\\

\noindent
\textbf {Example 1:} Analogously to the results that can be obtained by [8], we have the approximated solution for the FPBE when $0 <\alpha\leq 1$ using the initial condition $u(x,0)= g(x)$. The problem to be consider is 
\begin{equation}
\left\{
\begin{array}{l}
\displaystyle _c{\sf D}_t^{\alpha}u+u\frac{\partial u}{\partial x}-\frac{\partial^2 u}{\partial x^2}=0,\quad t>0,\quad 0<\alpha\leq{1}, \\\\
\displaystyle u(x,0)=g(x), \quad 0\leq x \leq 1.
\end{array} \right.
\end{equation}

\noindent
Using {\bf Corollary 1} we have $\lambda(t,\tau)=-1$ and  substitution  in the correction functional Eq.(15) we  obtain
 $$u_{n+1}(x,t)=u_n(x,t)-{\sf I}_t^{\alpha}\left[\dfrac{\partial^{\alpha}}{\partial t^{\alpha}}u_n-u_n(u_n)_x-(u_n)_{x x}\right].$$
 
\noindent
The correction functional can be written as
\begin{equation}
u_{n+1}(x,t)=u_n(x,t)-\frac{1}{\Gamma(\alpha)}\int_{0}^{t}(t-\tau)^{\alpha-1}[_c{\sf D}_t^\alpha u_n+u_n(u_n)_x-(u_n)_{x x}]{\mbox{d}}\tau.
\end{equation}

\noindent
Using Eq.(24) we obtain the following approximations

\begin{equation*}
 \displaystyle
 u_0= u(x,0)=g(x),
\end{equation*} 

 \noindent 
\begin{align*}
 \displaystyle
 u_1= g(x)-[g g^{\prime}-g^{\prime\prime}]{\displaystyle\frac{{t^{\alpha}}}{\Gamma(\alpha+1)}},
\end{align*} 
\begin{equation*}
 \displaystyle
 u_2= u_{1}+(2g(g^{\prime})^2+g^2g^{\prime\prime}-2g g^{(3)}-4g^{\prime}g^{\prime\prime}+g^{(4)}){{\frac{t^{2\alpha}}{\Gamma(1+2\alpha)}}}
\end{equation*} 
\noindent
\begin{equation*} 
\qquad\qquad\qquad\qquad\qquad\qquad -(gg^{\prime}-g^{\prime\prime})[(g^{\prime})^2+g g^{\prime\prime}-g^{(3)}]{{\frac{\Gamma(1+2\alpha)}{\Gamma^2(1+\alpha)}\frac{t^{3\alpha}}{\Gamma(1+3\alpha)}}}.
\end{equation*}

\noindent
Similar calculus are using for obtain others approximations. To close the section we prove another corollary.

\noindent
\begin{corol}
With $\lambda(t,\tau)=\displaystyle{\frac{(-1)^\alpha(\tau-t)^{\alpha-1}}{\Gamma(\alpha)}}$ and correction functional as
\begin{equation}
u_{n+1}(x,t)=u_n(x,t)+\frac{1}{\Gamma(\beta)}\int_{0}^{t}(t-\tau)^{\beta-1}\lambda(t,\tau)[_c{\sf D}_t^{\alpha}-(u_n)_{x x}-A(u_n)^p(u_n)_{x}]{\mbox{d}}\tau
\end{equation}
where $\beta=\alpha-[\alpha]$ and $[\alpha]$ is integer part of $\alpha$, then the iteration equation for $0<\alpha\leq 1$ and $1<\alpha\leq 2$, is\\
\begin{equation}
u_{n+1}=u_n-{\sf I}_t^{\alpha}[_c{\sf D}_t^{\alpha}-(u_n)_{x x}-A(u_n)^p(u_n)_{x}],
\end{equation}\\
and
\begin{equation}
u_{n+1}=u_n-(\alpha-1){\sf I}_t^{\alpha}[_c{\sf D}_t^{\alpha}u_n-(u_n)_{x x}-A(u_n)^p(u_n)_{x}].
\end{equation}

\noindent
respectively.
\end{corol}

\noindent
\textbf{Proof.}
By the {\bf Corollary 1}, $\lambda(t,\tau)=-1$ for $0<\alpha\leq{1}$ therefore $\beta=\alpha$, which imply

\begin{equation}
u_{n+1}(x,t)=u_n(x,t)-\frac{1}{\Gamma(\alpha)}\int_{0}^{t}(t-\tau)^{\alpha-1}[_c{\sf D}_t^\alpha u_n-A(u_n)^p(u_n)_x-(u_n)_{x x}]{\mbox{d}}\tau,
\end{equation}

\noindent
or in the following form

\begin{equation}
u_{n+1}(x,t)  =  u_n-{\sf I}_t^{\alpha}[_c{\sf D}_t^{\alpha}u_n-(u_n)_{x x}-A(u_n)^p(u_n)_x].
\end{equation}

\noindent
Now for $1<\alpha\leq 2$ therefore $\beta=\alpha-1$, by the {\bf Corollary 1}, we have $\lambda(t,\tau)=\tau-t$ and using the relation

\begin{align*}
\frac{1}{\Gamma(\alpha-1)}=\frac{-(\alpha-1)}{\Gamma(\alpha)},
\end{align*}

\noindent
we obtain
\begin{equation*}
{\displaystyle{u_{n+1}(x,t)  =  u_n+\frac{1}{\Gamma(\alpha-1)}\int_{0}^{t}(t-\tau)^{\alpha-1}[_c{\sf D}_t^{\alpha}u_n-(u_n)_{x x}-A(u_n)^p(u_n)_x]{\mbox {d}}\tau}},
\end{equation*}

\noindent
or in the following form

\begin{equation}
\noindent
{\displaystyle{u_{n+1}(x,t) =  u_n-(\alpha-1){\sf I}_t^{\alpha}[_c{\sf D}_t^{\alpha}u_n-(u_n)_{x x}-A(u_n)^p(u_n)_x]}}.
\end{equation}

\section{Approximate solutions for FBPE}

\noindent
There are some researchers that consider fractional Burgers' equation to model the diffusion behaviour of the flow through porous medium. In this section we will consider three examples of the fractional Burgers' equation, two of them with $0<\alpha\leq{1}$ and another one for the case $1<\alpha\leq{2}$, with its respective initial conditions. Here $u$ is the flow of velocity, the viscosity coefficient is consider equal to $-1$, and without  loss of generality we take $A=-1$ and $p=1$ in the Eq.(1). Note that, the viscosity coefficient  corresponds the second derivative in the Eq.(1). \\

\noindent
For those three examples, we consider an approximation of the solution in the following  series form

\begin{center}
$u(x,t)= u_{0}(x,t)+u_{1}(x,t)+u_{2}(x,t)+\cdots$
\end{center}

\noindent
To make the graphics, we stop the series in the three term. \\

\noindent
We mention  that the effect of the fractional derivative recover the memory effect associated with physical phenomena for   $0<\alpha\leq{1}$,  $1<\alpha\leq{2}$ \cite{Deng}. In particular with suitable initial conditions, when $A=0$, $\alpha=1$ and $\alpha=2$ in Eq.(1), we  recover the memory effect associated with a heat equation and wave equation, respectively. Also, with $\alpha=1$ in Eq.(1), we  obtain  the classical  Burgers' equation.

\noindent
To close this section we will present some examples with graphics to see better the effect involving  the parameter associated with the derivative.\\\\

\noindent
\textbf{Example 2:}
Consider the problem

\begin{equation}
\left\{
\begin{array}{l}
\displaystyle _c{\sf D}_t^{\alpha}u+u\frac{\partial u}{\partial x}-\frac{\partial^2 u}{\partial x^2}=0,\quad t>0,\quad 0<\alpha \leq {1}, \\
       
\\
\displaystyle u(x,0)=g(x)=\sin(\pi x) , \quad 0<x\leq{1}.
\end{array} \right.
\end{equation}\\

\noindent
Using Eq.(25) with $\lambda(t,\tau)=-1$ we obtain

\begin{equation*}
{\displaystyle{u_0(x,t)= \sin(\pi x)}},
\end{equation*}

\begin{equation*}
{\displaystyle{u_1(x,t)= \sin(\pi x)-\pi \sin(\pi x)[\cos(\pi x)+(\pi)^2]\frac{t^2}{\Gamma(\alpha+1)}}},
\end{equation*}

\begin{equation*}
\displaystyle
u_2(x,t)= u_{1}(x,t)+(2\pi^2 \sin(\pi x)\cos^2(\pi x)-\pi^2 \sin^4(\pi x)+2\pi^3 \sin(\pi x) \cos(\pi x)
\end{equation*}

\begin{equation*}
\qquad\qquad+4\pi^3 \cos(\pi x)\sin(\pi x)+\pi^4\sin^4(\pi x))
{\displaystyle{\frac{t^{2\alpha}}{\Gamma(1+2\alpha)}}}
\end{equation*}

\begin{equation*}
\qquad\qquad\qquad\qquad\qquad -\pi \sin(\pi x)\left[\cos(\pi x)+(\pi)^{2}\right]\pi^2 \cos(\pi x)-\pi^2 \sin^3(\pi x)
\end{equation*}

\begin{equation*}
\qquad\qquad \quad\qquad\qquad\qquad\qquad\qquad\qquad-\pi^4 \sin(\pi x))\displaystyle{\left(\frac{\Gamma(1+2\alpha)t^{3\alpha}}{\Gamma^2(1+\alpha)\Gamma(1+3\alpha)}\right)}.
\end{equation*}

\noindent
The graphics in Figure 1, with $\alpha=0.2$ and Figure 2 with $\alpha=1$ elucidate the approximation of $u(x,t)$.

\vspace{1 cm}

\begin{figure}[H]
    \centering
    \includegraphics [scale=0.6] {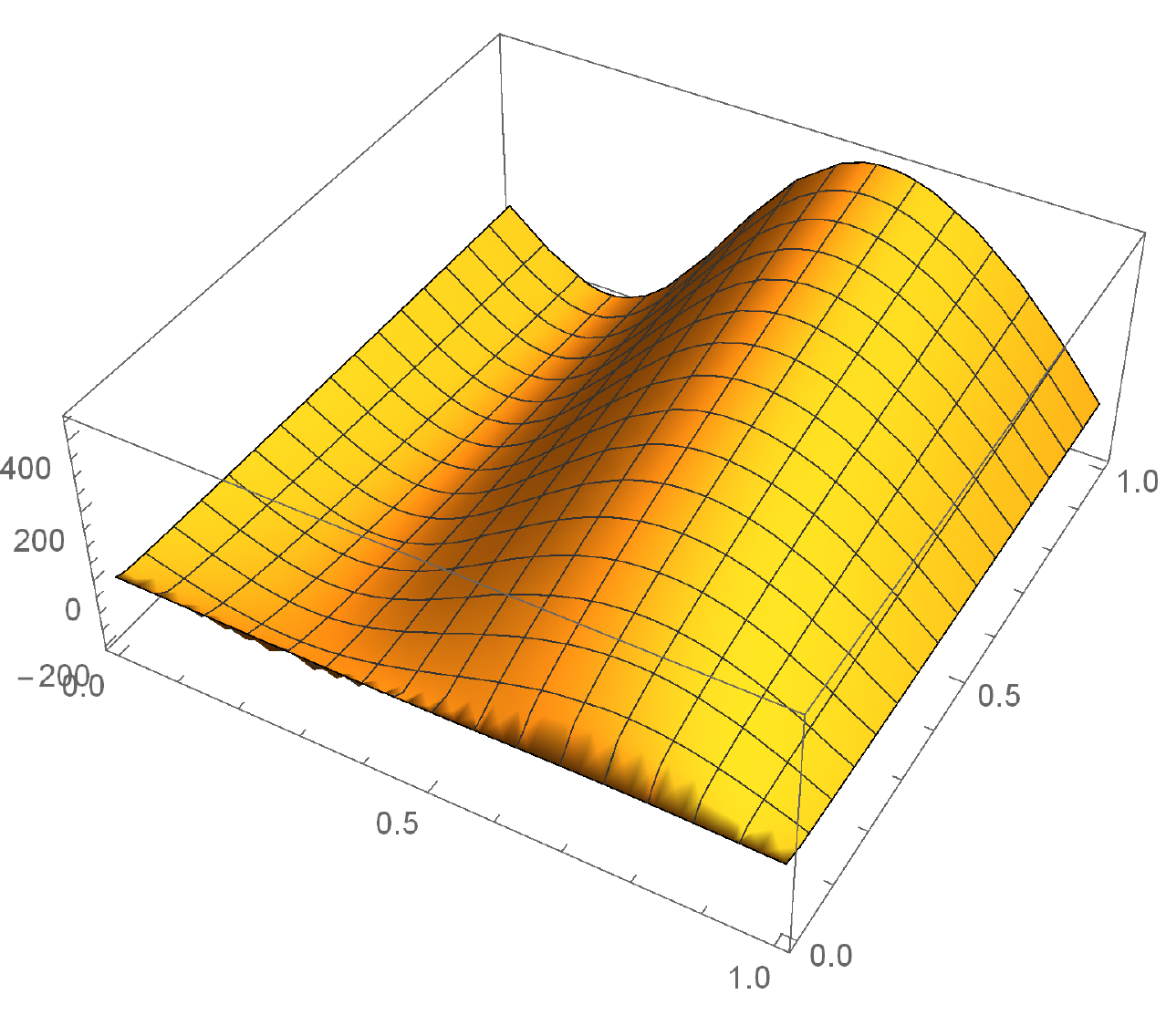}
    \caption{$u(x,0)=\sin(\pi x)$, with  $\alpha=0.2$.}
    \label{figRotulo}
  \end{figure}

\begin{figure}[H]
    \centering
    \includegraphics [scale=0.6] {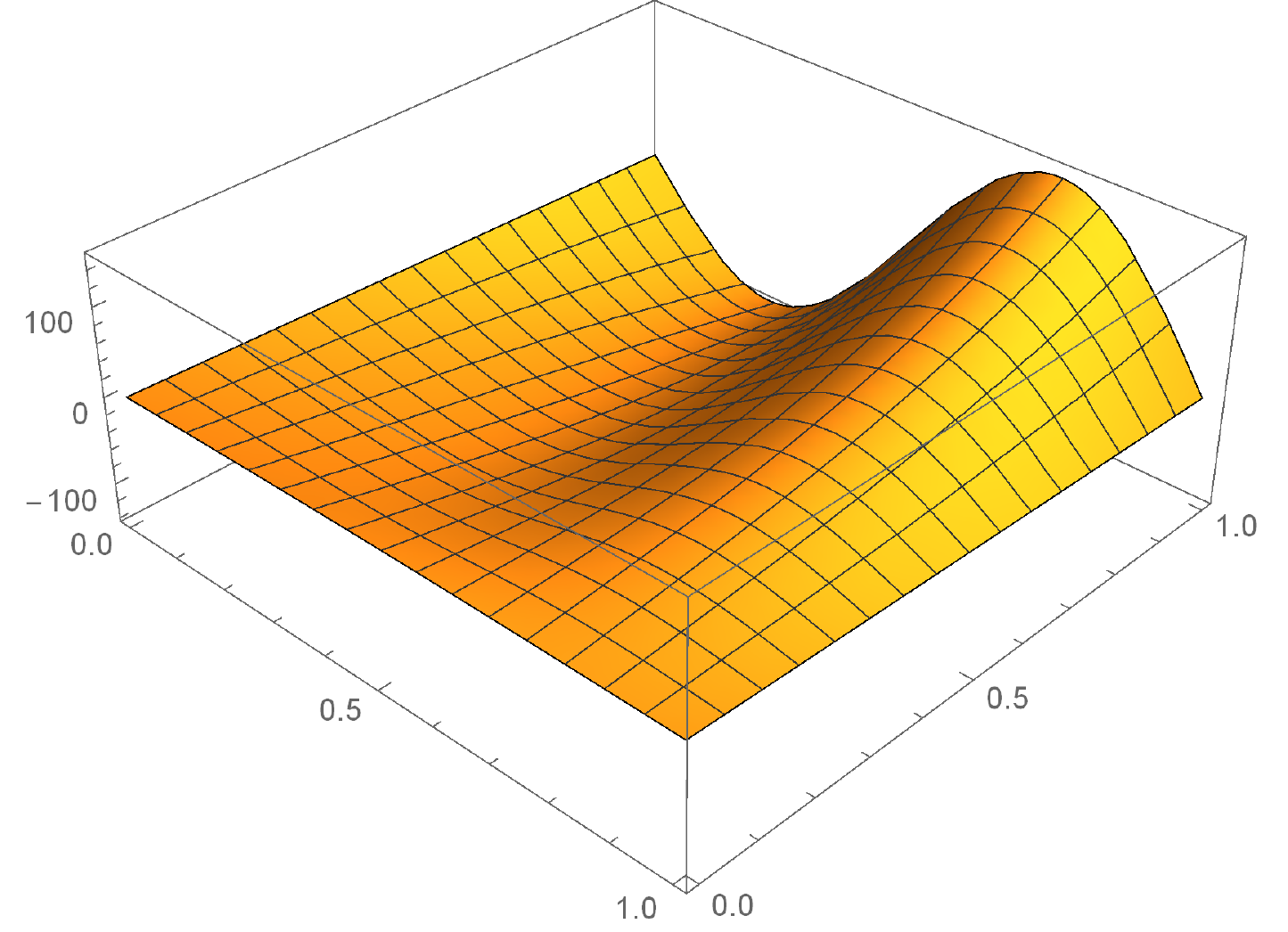}
    \caption{$u(x,0)=\sin(\pi x)$, with  $\alpha=1$.}
    \label{figRotulo}
  \end{figure}

\noindent
\textbf{Example 3:} Consider the  Example 2 with the initial condition given by $ u(x,0)=g(x)={\displaystyle{\frac{e^x}{1+e^x}}}$.
Using Eq.(25) with $\lambda(t,\tau)=-1$ we get

\begin{equation*}
\displaystyle{u_{0}(x,t)  =  \frac{e^x}{1+e^x}},
\end{equation*}

\begin{equation*}
{\displaystyle{u_1(x,t) =  \frac{e^x}{1+e^x}-\left(\frac{e^{x}(e^{x}+e^{2x}-1)}{(1+e^{x})^3}\right) \left(\frac{t^\alpha}{\Gamma(\alpha+1)}\right)}},
\end{equation*}
\\\\

\noindent
\begin{equation*}
{\displaystyle{u_2(x,t) =\frac{e^x}{1+e^x}- \left(\frac{e^{x}(e^{x}+e^{2x}-1)}{(1+e^{x})^3}\right)   \left(\frac{t^\alpha}{\Gamma(\alpha+1)}\right)}}
\end{equation*}

\begin{equation*}
\qquad\qquad\qquad +\left(\displaystyle{\frac{e^{x}(1-12e^{x}-29e^{2x}+4e^{3x}+37e^{4x}+50e^{5x}+11e^{6x}+2e^{7x})}{(1+e^{x})^8}} \right)
\end{equation*}

\begin{equation*}
\qquad\qquad\qquad\qquad\times\left(\frac{t^\alpha}{\Gamma(1+2\alpha)}\right)-
\left(\displaystyle{\frac{e^{x}(1-4e^{x}-10e^{2x}+6e^{3x}+2e^{4x}+e^{5x})}{(1+e^{x})^6} }\right)
\end{equation*}

\begin{equation*}
\qquad\qquad\qquad\qquad\qquad\qquad\qquad\qquad\qquad\qquad\qquad\quad\times\left(\displaystyle{\frac{\Gamma(1+2\alpha)}{\Gamma^2(1+\alpha)}\frac{t^{3\alpha}}{\Gamma(1+3\alpha)}}\right).
\end{equation*}

\noindent
Also here, the graphics in Figure 3, with $\alpha=0.2$  and Figure 4, with $\alpha=1$, elucidate the approximation of $u(x,t)$.

\begin{figure}[H]
    \centering
    \includegraphics [scale=0.5] {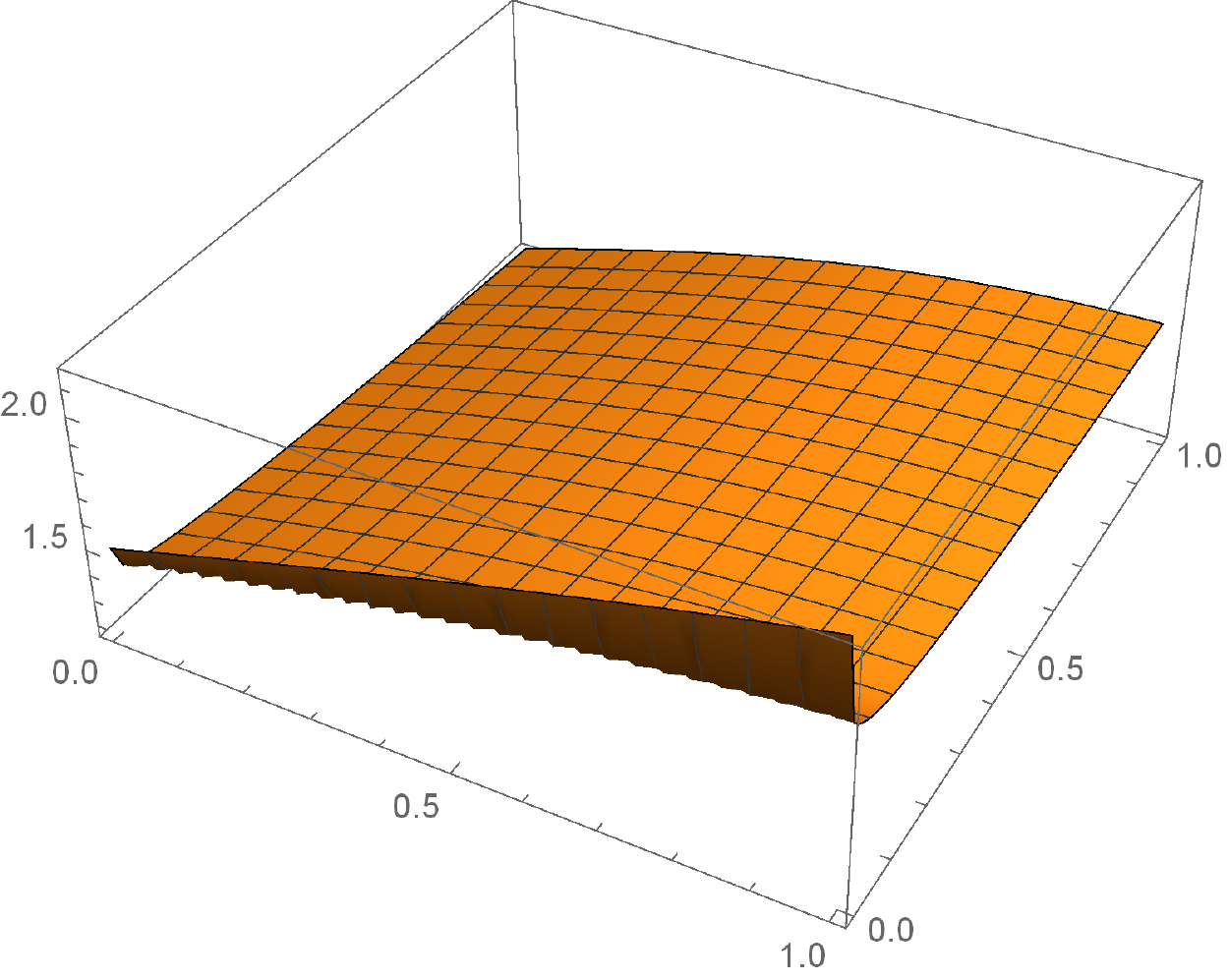}
    \caption{$u(x,0)=\displaystyle \frac{e^x}{1+e^x}$, with $\alpha=0.2$.}
    \label{figRotulo}
\end{figure}
\begin{figure}[H]
    \centering
    \includegraphics [scale=0.5] {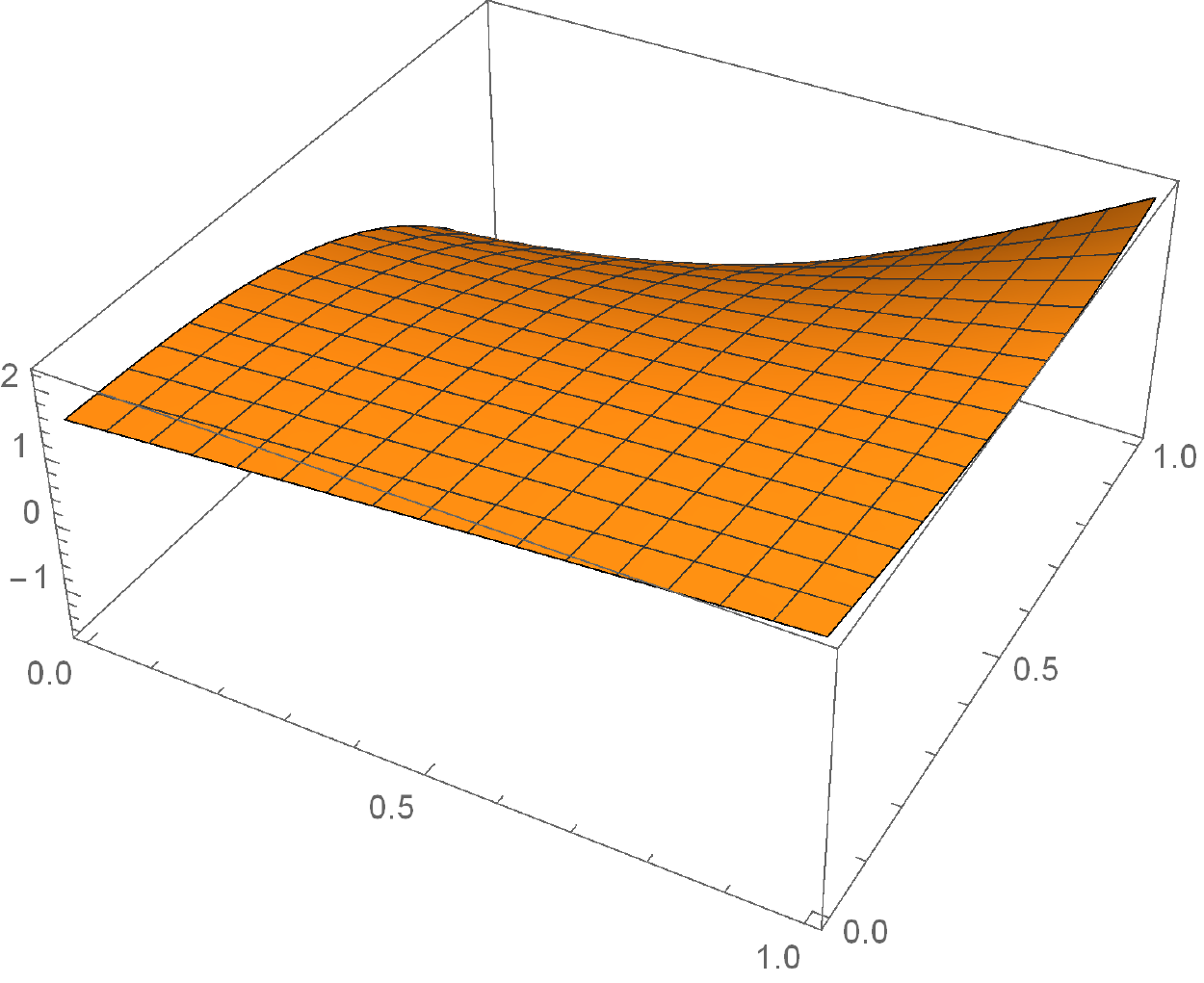}
    \caption{$u(x,0)=\displaystyle \frac{e^x}{1+e^x}$, with $\alpha=1$.}
    \label{figRotulo}
\end{figure}

\noindent
\textbf{Example 4:}
Consider the problem

\begin{equation*}
\left\{
\begin{array}{l}
\displaystyle _c{\sf D}_t^{\alpha}u+u\frac{\partial u}{\partial x}-\frac{\partial^2 u}{\partial x^2}=0,\quad t>0,\quad 1< \alpha \leq 2, \\
       \\
\displaystyle u(x,0)=\sin(\pi x),\qquad u_t(x,0)=0,\quad 0\leq x \leq 1.
\end{array} \right.
\end{equation*}\\

\noindent
In this case, using Eq.(25) with $\lambda(t,\tau)=\tau-t$ we  obtain 

\begin{equation*}
\noindent
{\displaystyle{u_0(x,t)=\sin(\pi x)}},
\end{equation*}
\begin{equation*}
\noindent
{\displaystyle{u_1(x,t)  =  \sin(\pi x)-\alpha(\alpha-1)[\pi \sin(\pi x)\cos(\pi x)-\pi^2 \sin(x)]\left[\frac{t^\alpha}{\Gamma(\alpha+1)}\right]}},
\noindent
\end{equation*}
\begin{equation*}
{\displaystyle{u_2(x,t)  =  \sin(\pi x)-\alpha(\alpha-1)[\pi \sin(\pi x)\cos(\pi x)-\pi^2 \sin(x)]\left[\frac{t^\alpha}{\Gamma(\alpha+1)}\right]}}
\end{equation*}
\begin{equation*}
\qquad\qquad+\alpha(\alpha-1)[\sin(\pi x)(2\pi^2 \cos^2(\pi x)-\pi^2 \sin^3(\pi x)+6\pi^3 \cos(\pi x)
\end{equation*}
\begin{equation*}
\qquad\qquad\qquad+\pi^4 \sin^3(\pi x)]\left[\frac{t^2\alpha}{\Gamma(1+2\alpha)}\right]-\alpha(\alpha-1)(\pi \sin(\pi x))(\cos(\pi x)+x^2)
\end{equation*}
\begin{equation*}
\qquad\qquad\qquad\quad\times(\pi^2 \cos(\pi x)-\pi^2 \sin^3(\pi x)-\pi^4 \sin(\pi x)\left[\frac{\Gamma(1+2\alpha)t^3\alpha}{\Gamma^2(1+\alpha)\Gamma(1+3\alpha)}\right].
\end{equation*}

\noindent
The graphics in Figure 5, with $\alpha=1.2$ and Figure 6, with $\alpha=2$ one can see the evolution of the approximation of $u(x,t)$.

\begin{figure}[H]
    \centering
    \includegraphics [scale=0.5] {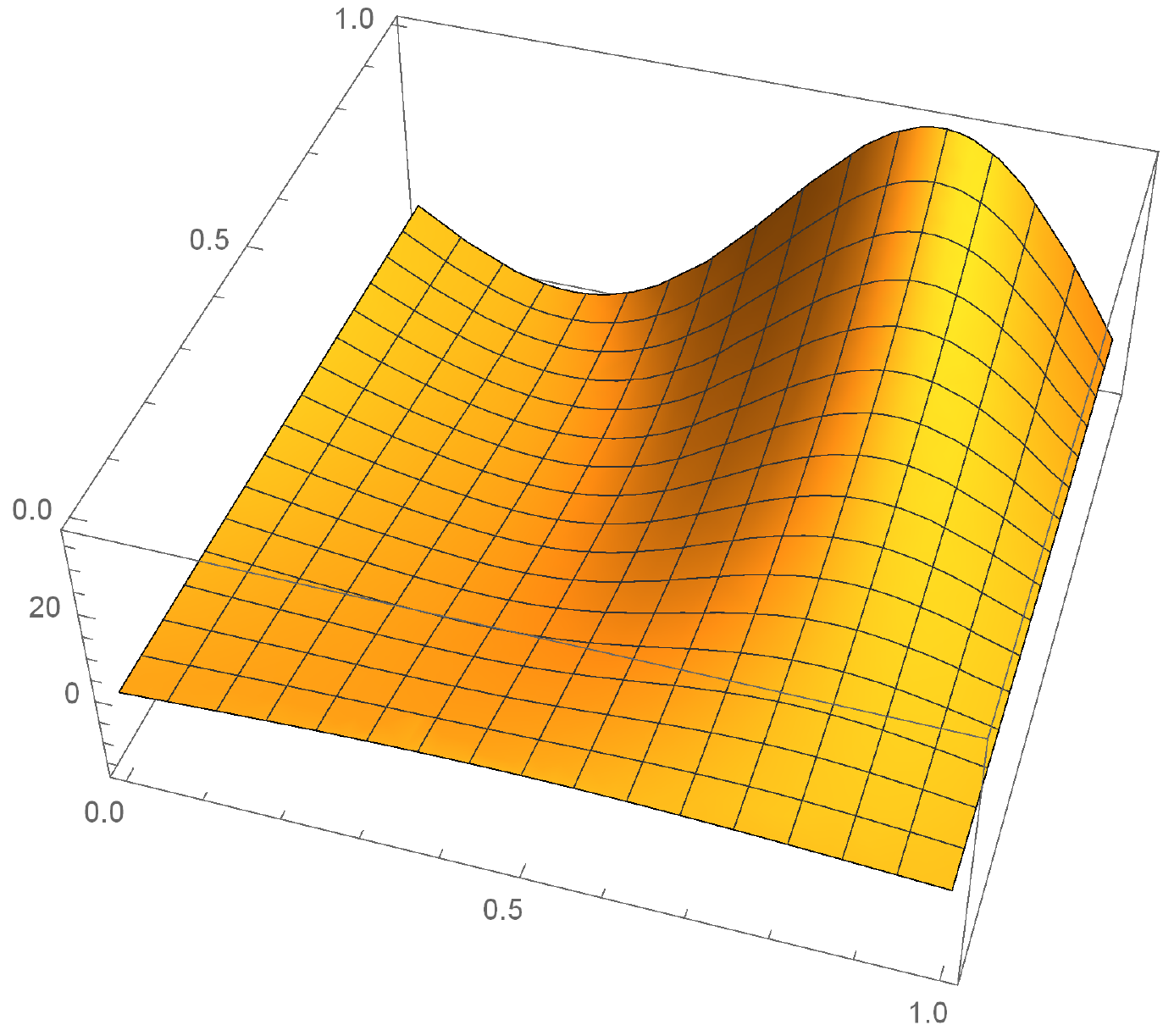}
    \caption{$u(x,0)=\sin(\pi x),u_{t}(x,0)=0$, with $\alpha=1.2$.}
   \label{figRotulo}
  \end{figure}
  
 \begin{figure}[H]
    \centering
    \includegraphics [scale=0.5] {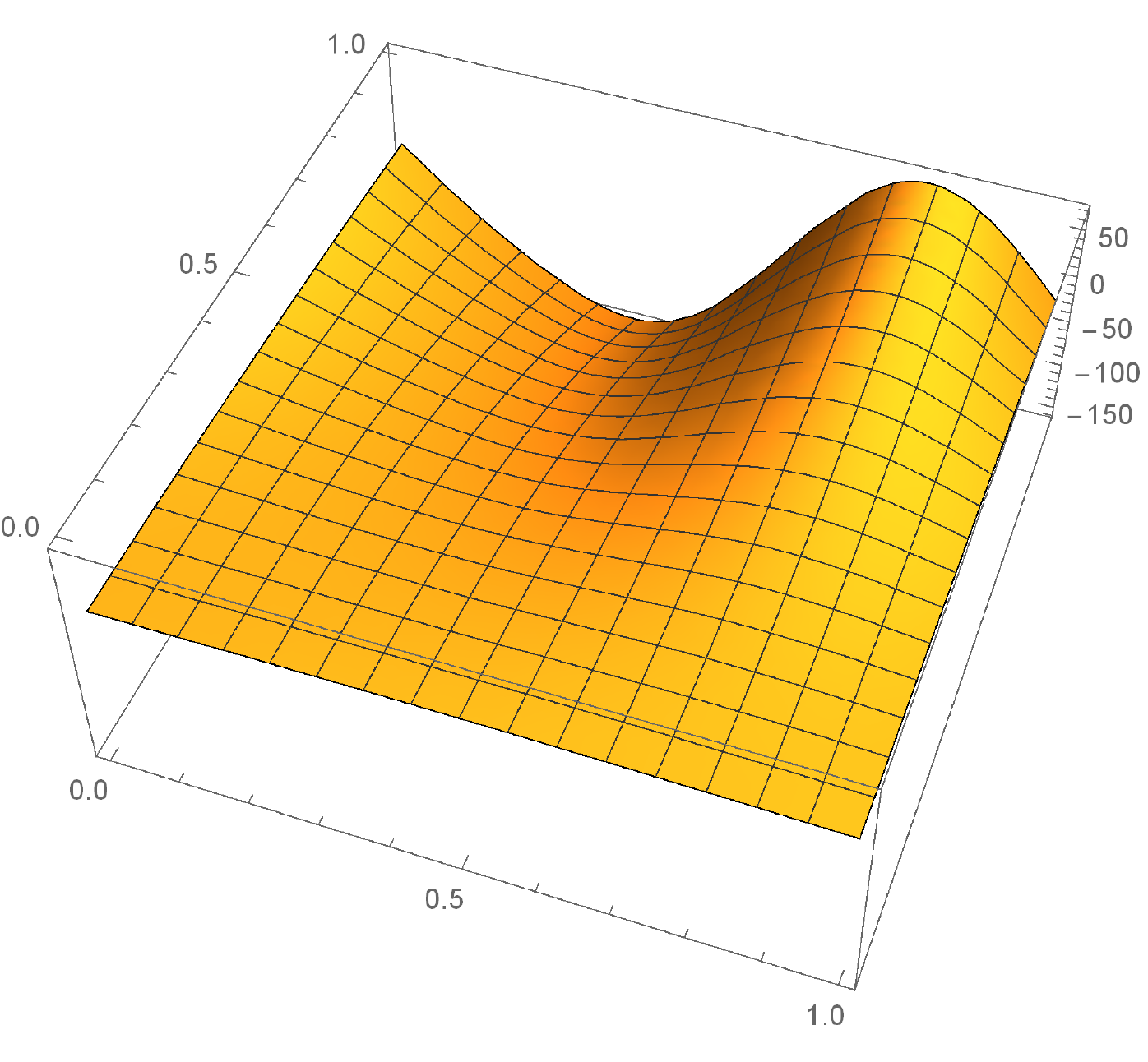}
    \caption{$u(x,0)=\sin(\pi x),u_{t}(x,0)=0$, with $\alpha=2$.}
   \label{figRotulo}
  \end{figure}

\section{Conclusions}

The FC is very useful in the recuperation of the memory of phenomena, by used of fractional derivative in  time variable. News Lagrange multipliers for FPBE are  identified by Laplace transform and particular cases are recovered. Using this multipliers and the VIM we obtained approximations of the solutions for FPBE taking three terms only. Then, we conclude that the VIM is a powerful and efficient technical to approximate solutions to FPBE.


\end{document}